\documentclass[aps,prl,twocolumn,tightenlines,superscriptaddress,showpacs,byrevtex]{revtex4-1}

\usepackage{epsfig,floatflt}
\usepackage{graphicx} 
\usepackage{dcolumn}  
\usepackage{amssymb,amsmath,indentfirst,enumerate}

\graphicspath{{ps}}

\newcommand{\UFS}{\ensuremath{\Upsilon(10860)}}
\newcommand{\pp}{\ensuremath{\pi^+\pi^-}}

\newcommand{\ee}{\ensuremath{e^+e^-}}
\newcommand{\Zbl}{\ensuremath{Z_b(10610)}}
\newcommand{\Zbh}{\ensuremath{Z_b(10650)}}

\newcommand{\bbpi}{\ensuremath{B^{(*)}B^{(*)}\pi}}
\newcommand{\bbstpi}{\ensuremath{BB^*\pi}}
\newcommand{\bstbstpi}{\ensuremath{B^{*}B^{*}\pi}}

\begin{document}

\title{\quad\\[0.5cm]
Study of $\ee\to B^{(*)}\bar{B}^{(*)}\pi^\pm$ at $\sqrt{s}=10.866$~GeV}


\noaffiliation
\affiliation{University of the Basque Country UPV/EHU, 48080 Bilbao}
\affiliation{Beihang University, Beijing 100191}
\affiliation{Budker Institute of Nuclear Physics SB RAS, Novosibirsk 630090}
\affiliation{Faculty of Mathematics and Physics, Charles University, 121 16 Prague}
\affiliation{Chonnam National University, Kwangju 660-701}
\affiliation{University of Cincinnati, Cincinnati, Ohio 45221}
\affiliation{Deutsches Elektronen--Synchrotron, 22607 Hamburg}
\affiliation{University of Florida, Gainesville, Florida 32611}
\affiliation{Justus-Liebig-Universit\"at Gie\ss{}en, 35392 Gie\ss{}en}
\affiliation{SOKENDAI (The Graduate University for Advanced Studies), Hayama 240-0193}
\affiliation{Hanyang University, Seoul 133-791}
\affiliation{University of Hawaii, Honolulu, Hawaii 96822}
\affiliation{High Energy Accelerator Research Organization (KEK), Tsukuba 305-0801}
\affiliation{IKERBASQUE, Basque Foundation for Science, 48013 Bilbao}
\affiliation{Indian Institute of Technology Bhubaneswar, Satya Nagar 751007}
\affiliation{Indian Institute of Technology Guwahati, Assam 781039}
\affiliation{Indian Institute of Technology Madras, Chennai 600036}
\affiliation{Indiana University, Bloomington, Indiana 47408}
\affiliation{Institute of High Energy Physics, Chinese Academy of Sciences, Beijing 100049}
\affiliation{Institute of High Energy Physics, Vienna 1050}
\affiliation{INFN - Sezione di Torino, 10125 Torino}
\affiliation{Institute for Theoretical and Experimental Physics, Moscow 117218}
\affiliation{J. Stefan Institute, 1000 Ljubljana}
\affiliation{Kanagawa University, Yokohama 221-8686}
\affiliation{Institut f\"ur Experimentelle Kernphysik, Karlsruher Institut f\"ur Technologie, 76131 Karlsruhe}
\affiliation{Kennesaw State University, Kennesaw GA 30144}
\affiliation{King Abdulaziz City for Science and Technology, Riyadh 11442}
\affiliation{Korea Institute of Science and Technology Information, Daejeon 305-806}
\affiliation{Korea University, Seoul 136-713}
\affiliation{Kyungpook National University, Daegu 702-701}
\affiliation{\'Ecole Polytechnique F\'ed\'erale de Lausanne (EPFL), Lausanne 1015}
\affiliation{Faculty of Mathematics and Physics, University of Ljubljana, 1000 Ljubljana}
\affiliation{Luther College, Decorah, Iowa 52101}
\affiliation{University of Maribor, 2000 Maribor}
\affiliation{Max-Planck-Institut f\"ur Physik, 80805 M\"unchen}
\affiliation{School of Physics, University of Melbourne, Victoria 3010}
\affiliation{Moscow Physical Engineering Institute, Moscow 115409}
\affiliation{Moscow Institute of Physics and Technology, Moscow Region 141700}
\affiliation{Graduate School of Science, Nagoya University, Nagoya 464-8602}
\affiliation{Kobayashi-Maskawa Institute, Nagoya University, Nagoya 464-8602}
\affiliation{Nara Women's University, Nara 630-8506}
\affiliation{National Central University, Chung-li 32054}
\affiliation{National United University, Miao Li 36003}
\affiliation{Department of Physics, National Taiwan University, Taipei 10617}
\affiliation{H. Niewodniczanski Institute of Nuclear Physics, Krakow 31-342}
\affiliation{Niigata University, Niigata 950-2181}
\affiliation{Novosibirsk State University, Novosibirsk 630090}
\affiliation{Osaka City University, Osaka 558-8585}
\affiliation{Pacific Northwest National Laboratory, Richland, Washington 99352}
\affiliation{University of Pittsburgh, Pittsburgh, Pennsylvania 15260}
\affiliation{University of Science and Technology of China, Hefei 230026}
\affiliation{Seoul National University, Seoul 151-742}
\affiliation{Soongsil University, Seoul 156-743}
\affiliation{University of South Carolina, Columbia, South Carolina 29208}
\affiliation{Sungkyunkwan University, Suwon 440-746}
\affiliation{School of Physics, University of Sydney, NSW 2006}
\affiliation{Department of Physics, Faculty of Science, University of Tabuk, Tabuk 71451}
\affiliation{Tata Institute of Fundamental Research, Mumbai 400005}
\affiliation{Excellence Cluster Universe, Technische Universit\"at M\"unchen, 85748 Garching}
\affiliation{Tohoku University, Sendai 980-8578}
\affiliation{Earthquake Research Institute, University of Tokyo, Tokyo 113-0032}
\affiliation{Department of Physics, University of Tokyo, Tokyo 113-0033}
\affiliation{Tokyo Institute of Technology, Tokyo 152-8550}
\affiliation{Tokyo Metropolitan University, Tokyo 192-0397}
\affiliation{University of Torino, 10124 Torino}
\affiliation{CNP, Virginia Polytechnic Institute and State University, Blacksburg, Virginia 24061}
\affiliation{Wayne State University, Detroit, Michigan 48202}
\affiliation{Yamagata University, Yamagata 990-8560}
\affiliation{Yonsei University, Seoul 120-749}
  \author{A.~Garmash}\affiliation{Budker Institute of Nuclear Physics SB RAS, Novosibirsk 630090}\affiliation{Novosibirsk State University, Novosibirsk 630090} 
  \author{A.~Abdesselam}\affiliation{Department of Physics, Faculty of Science, University of Tabuk, Tabuk 71451} 
  \author{I.~Adachi}\affiliation{High Energy Accelerator Research Organization (KEK), Tsukuba 305-0801}\affiliation{SOKENDAI (The Graduate University for Advanced Studies), Hayama 240-0193} 
  \author{H.~Aihara}\affiliation{Department of Physics, University of Tokyo, Tokyo 113-0033} 
  \author{D.~M.~Asner}\affiliation{Pacific Northwest National Laboratory, Richland, Washington 99352} 
  \author{T.~Aushev}\affiliation{Moscow Institute of Physics and Technology, Moscow Region 141700}\affiliation{Institute for Theoretical and Experimental Physics, Moscow 117218} 
  \author{R.~Ayad}\affiliation{Department of Physics, Faculty of Science, University of Tabuk, Tabuk 71451} 
  \author{T.~Aziz}\affiliation{Tata Institute of Fundamental Research, Mumbai 400005} 
  \author{V.~Babu}\affiliation{Tata Institute of Fundamental Research, Mumbai 400005} 
  \author{I.~Badhrees}\affiliation{Department of Physics, Faculty of Science, University of Tabuk, Tabuk 71451}\affiliation{King Abdulaziz City for Science and Technology, Riyadh 11442} 
% ...
\collaboration{The Belle Collaboration}


\begin{abstract}

We report the analysis of the three-body 
$\ee\to B\bar{B}\pi$, $B\bar{B}^*\pi$, and $B^*\bar{B}^*\pi$ processes,
including the first observation of the 
$Z^{\pm}_b(10610)\to [B\bar{B}^*+{\rm c.c.}]^{\pm}$ and 
$Z^{\pm}_b(10650)\to [B^*\bar{B}^*]^\pm$ 
transitions. We measure visible cross sections for the three-body 
production of
$\sigma_{\rm vis}(\ee\to [B\bar{B}^*+{\rm c.c.}]^{\pm}\pi^{\mp})=
(11.2\pm1.0(stat.)\pm1.2(syst.))$~pb and
$\sigma_{\rm vis}(\ee\to [B^*\bar{B}^*]^{\pm}\pi^{\mp})=
(5.61\pm0.73(stat.)\pm0.66(syst.))$~pb and set a 90\% C.L.\ upper limit of
$\sigma_{\rm vis}(\ee\to [B\bar{B}]^{\pm}\pi^{\mp})<2.1$~pb. 
The results are based on a $121.4$~fb$^{-1}$ data sample collected with
the Belle detector at a center-of-mass energy near the $\UFS$ peak. 

\end{abstract}

\maketitle

Two new charged bottomonium-like resonances, $Z_b(10610)$ and 
$Z_b(10650)$, have been observed recently by the Belle Collaboration 
in $\ee\to\Upsilon(n{\rm S})\pi^+\pi^-$, $n=1,2,3$ and 
$\ee\to h_b(m{\rm P})\pi^+\pi^-$, $m=1,2$~\cite{y5s2ypp,y5s2ypp2}. 
Analysis of the quark composition of the initial and final 
states reveals that these hadronic objects have 
an exotic nature: $Z_b$ should be comprised of (at least)
four quarks including a $b\bar{b}$ pair. Several 
models~\cite{zbmodels} have been proposed to describe the internal 
structure of these states. In Ref.~\cite{molec}, it was suggested that 
$Z_b(10610)$ and $Z_b(10650)$ states might be loosely bound 
$B\bar{B}^*$ and $B^*\bar{B}^*$ systems, respectively. If so, 
it is natural to expect the $Z_b$ states to decay to 
final states with $B^{(*)}$ mesons at substantial rates.

Evidence for the three-body $\UFS\to B\bar{B}^*\pi$ decay has 
been reported previously by Belle, based on a data sample of 
$23.6$~fb$^{-1}$~\cite{drutskoj}. In this analysis, we use a data 
sample with an integrated luminosity of $121.4$~fb$^{-1}$ collected 
near the peak of the $\UFS$ resonance ($\sqrt{s}=10.866$~GeV) with 
the Belle detector~\cite{Belle} at the KEKB asymmetric-energy $\ee$ 
collider~\cite{KEKB}. 
Note that we reconstruct only three-body $B^{(*)}\bar{B}^{(*)}\pi$ 
combinations with a charged primary pion. For brevity, we adopt the 
following notations: the set of $B^+\bar{B}^0\pi^-$ and $B^-B^0\pi^+$ 
final states is referred to as $BB\pi$; 
the set of $B^+\bar{B}^{*0}\pi^-$, $B^-B^{*0}\pi^+$, 
$B^0B^{*-}\pi^+$ and $\bar{B}^0B^{*+}\pi^-$ final states is referred 
to as $BB^*\pi$; and the set of $B^{*+}\bar{B}^{*0}\pi^-$ and 
$B^{*-}B^{*0}\pi^+$ final states is denoted as $B^*B^*\pi$.
The inclusion of the charge conjugate mode is implied throughout 
this report.

We use Monte Carlo (MC)
events generated with EvtGen~\cite{EVTGEN} and then processed through 
a detailed detector simulation implemented in GEANT3~\cite{GEANT}. 
Final-state radiation from charged particles is simulated during event 
generation using PHOTOS~\cite{PHOTOS}.
The simulated samples for $\ee\to q\bar{q}$ ($q=u$, $d$, $s$, $c$, 
or $b$) are equivalent to six times the integrated 
luminosity of the data and are used to develop criteria to separate signal 
events from backgrounds, identify types of background events, determine 
the reconstruction efficiency and parameterize the distributions needed 
for the extraction of the signal decays.


$B$ mesons are reconstructed in the following decay channels:
$B^+\to J/\psi K^{(*)+}$, $B^+\to \bar{D}^{(*)0}\pi^+$, 
$B^0\to J/\psi K^{(*)0}$, $B^0\to D^{(*)-}\pi^+$ (eighteen in total).
Charged-track candidates are required to be consistent with 
origination from the interaction point (IP). A likelihood ratio 
for a given track to be a $\pi$, $K$, or $p$ is obtained by utilizing 
energy-loss measurements in the (CDC), light yield measurements from 
an array of aerogel threshold Cherenkov counters, and time-of-flight 
information from a barrel-like arrangement of time-of-flight scintillation 
counters. Photons are detected with an electromagnetic calorimeter (ECL) 
and are required to have energies in the laboratory frame of at least 
50 (100)~MeV in the ECL barrel (endcaps) and not be associated with 
charged tracks. 

Neutral pion candidates are reconstructed using photon pairs with an 
invariant mass between 120 and 150 MeV/$c^2$. Neutral kaon candidates 
are reconstructed using pairs of oppositely-charged pions with an 
invariant mass within 15~MeV/$c^2$ of the nominal $K^0$ mass. The 
direction of the $K^0$ candidate momentum vector is required to be 
consistent with the direction of its vertex displecement relative to 
the IP. The $K^{*0}$ ($K^{*+}$) 
is reconstructed in the $K^+\pi^-$ ($K^0\pi^+$) final state, the 
invariant mass of the $K^{*}$ candidate is required to be within 
150~MeV/$c^2$ of the nominal $K^{*}$ mass~\cite{PDG}. The invariant 
mass of a $J/\psi\to \ell^+\ell^-$ candidate is required to be within 
30 (50)~MeV/$c^2$ for $\ell=e$ ($\mu$), of the nominal $J/\psi$ mass. 
Neutral (charged) $D$ mesons are reconstructed in the $K^-\pi^+$, 
$K^-\pi^+\pi^0$, and $K^-\pi^-\pi^+\pi^+$ ($K^-\pi^+\pi^+$) modes. 
To identify $D^{*}$ candidates, we require 
$|M(D\pi)-M(D)-\Delta m_{D^*}|<3$~MeV/$c^2$, where $M(D\pi)$ and 
$M(D)$ are the reconstructed masses of the $D^*$ and $D$ candidates, 
respectively, and $\Delta m_{D^*}=m_{D^*}-m_D$ is the difference between 
the nominal $D^*$ and $D$ masses. The mass windows for narrow states
quoted above correspond to a $\pm2.5\sigma$ requirement.

\begin{figure}[t]
  \includegraphics[width=0.245\textwidth]{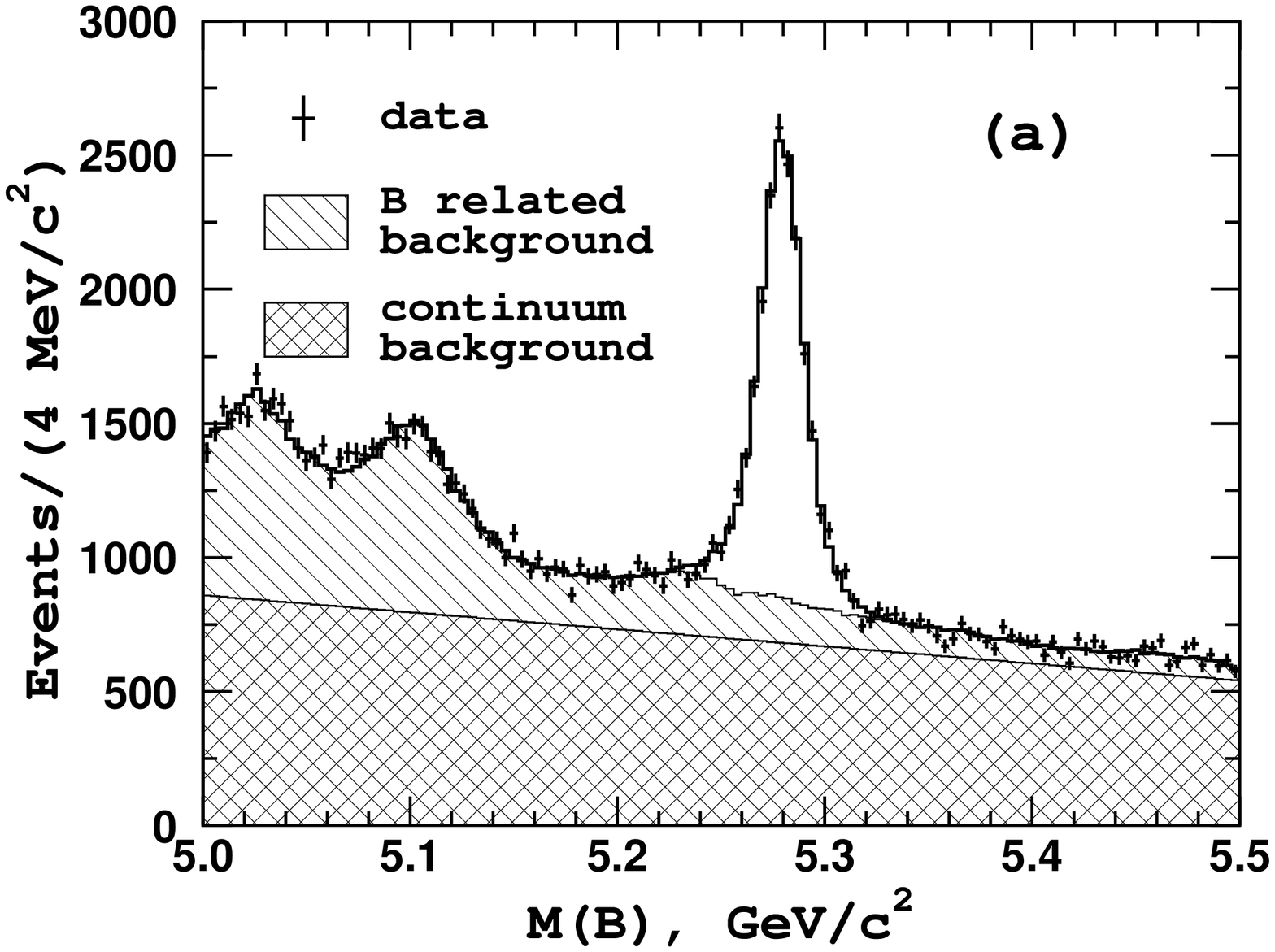} \hfill
  \hspace*{-5mm}
  \includegraphics[width=0.245\textwidth]{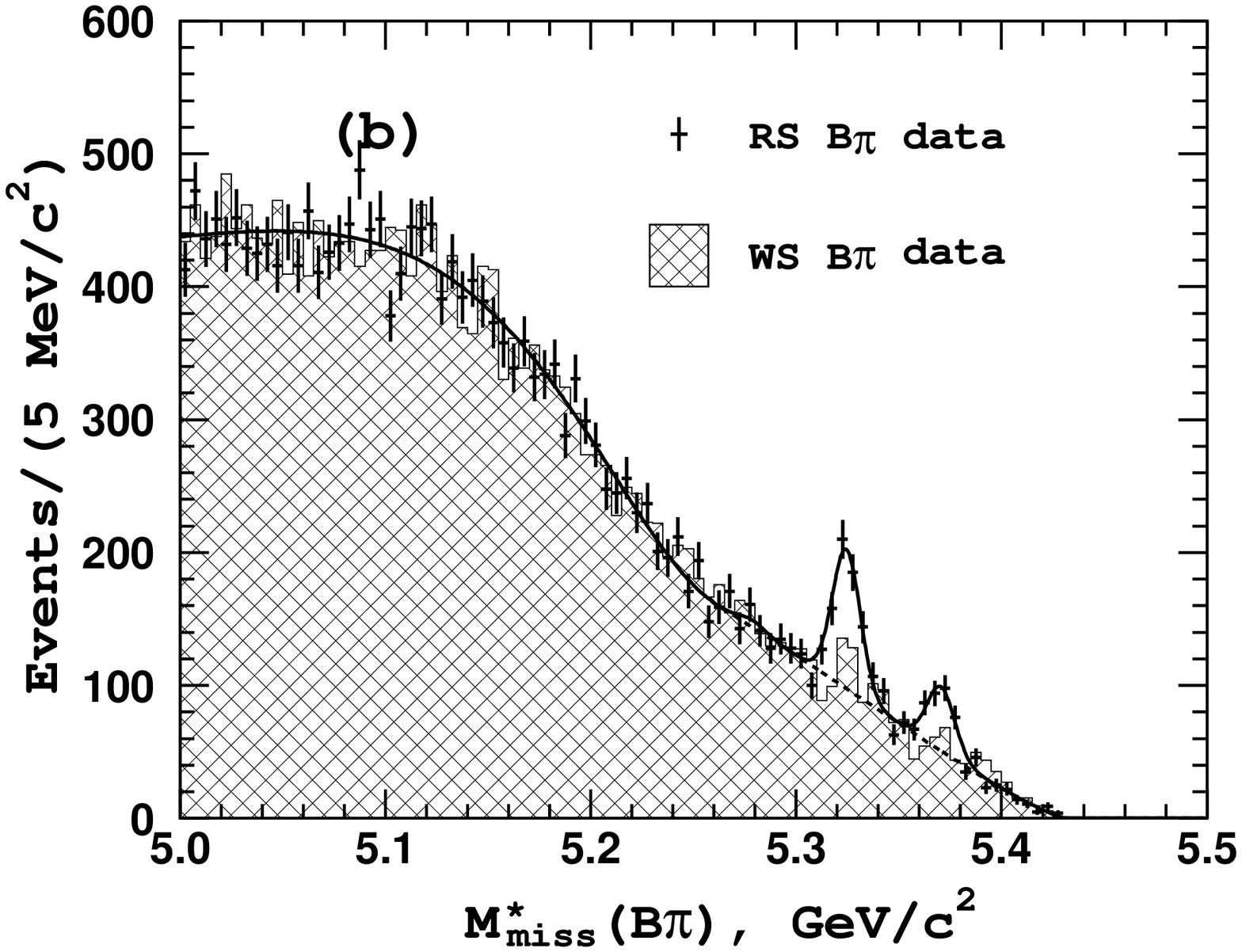}
  \caption{The (a) invariant mass and 
           (b) $M^*_{\rm miss}(B\pi)$ 
           distribution for $B$ candidates in the $B$ signal region.
           Points with error bars represent the data. The open histogram
           in (a) shows the result of the fit to data. The solid line 
           in (b) shows the result of the fit to the RS $B\pi$ data; 
           the dashed line represents the background level.}
  \label{fig:select}
\end{figure}

The dominant background comes from $e^+e^-\to c\bar{c}$ continuum events, 
where true $D$ mesons produced in $e^+e^-$ annihilation are combined with 
random particles to form a $B$ candidate. This type of background is
suppressed using variables that characterize the event topology. Since the
momenta of the two $B$ mesons produced from a three-body $\ee\to\bbpi$ decay 
are low in the c.m.\ frame (below 0.9~GeV/$c$), their decay products are 
essentially uncorrelated so that the event tends to be spherical. In contrast, 
hadrons from continuum events tend to exhibit a back-to-back jet 
structure. We use $\theta_{\rm thr}$, the angle between the thrust axis of 
the $B$ candidate and that of the rest of the event, to discriminate 
between the two cases. The distribution of 
$|\cos\theta_{\rm thr}|$ is strongly peaked near $|\cos\theta_{\rm thr}|=1.0$
for $c\bar{c}$ events and is nearly flat for $\bbpi$ events. We require
$|\cos\theta_{\rm thr}|<0.80$ for the $B\to D^{(*)}\pi$ final states; this
eliminates about 81\% of the continuum background and retains 73\% of the
signal events.

We identify $B$ candidates by their reconstructed invariant mass $M(B)$ 
and momentum $P(B)$ in the center-of-mass (c.m.) frame. 
We require $P(B)<1.35$~GeV/$c$ to retain $B$ mesons produced in both
two-body and multibody processes.
The $M(B)$ 
distribution for $B$ candidates is shown in 
Fig.~\ref{fig:select}(a).
We perform a binned maximum likelihood fit of the $M(B)$ distribution 
to the sum of a signal component parameterized by a Gaussian function 
and two background components: one related to other decay modes of $B$ 
mesons and one due to continuum $e^+e^-\to q\bar{q}$ processes, where 
$q=u,d,s,c$. The shape of the $B$-related background is determined from 
a large sample of generic MC; the shape of the $q\bar{q}$ background is 
parameterized with a linear function. The parameters of the signal 
Gaussian, the normalization of the $B$-related background and the 
parameters of the $q\bar{q}$ background float in the fit. 
We find $12263\pm168$ fully reconstructed $B$ mesons. The $B$ signal 
region is defined by requiring $M(B)$ to be within 30 to 40~MeV/$c^2$ 
(depending on the $B$ decay mode) of the nominal $B$ mass.

Reconstructed $B^+$ or $\bar{B}^0$ candidates are combined with 
$\pi^-$'s --- the right-sign (RS) combination --- and the missing 
mass, $M_{\rm miss}(B\pi)$, is calculated as 
$M_{\rm miss}(B\pi) = \sqrt{(\sqrt{s}-E_{B\pi})^2/c^4-P^2_{B\pi}/c^2}$, 
where $E_{B\pi}$ and $P_{B\pi}$ are the measured energy and 
momentum of the reconstructed $B\pi$ combination.
Signal $\ee\to \bbstpi$ events produce a narrow peak in 
the $M_{\rm miss}(B\pi)$ spectrum around the nominal $B^*$ mass while 
$\ee\to \bstbstpi$ events produce a peak at $m_{B^*}+\Delta m_{B^*}$,
where $\Delta m_{B^*}=m_{B^*}-m_{B}$, due to the missed photon from the
$B^*\to B\gamma$ decay. It is important to note here that, according 
to signal MC, $\bbstpi$ events, where the reconstructed $B$ is the 
one from the $B^*$, produce a peak in the $M_{\rm miss}(B\pi)$ 
distribution at virtually the same position as $\bbstpi$ events, 
where the reconstructed $B$ is the primary one. To remove the 
correlation between $M_{\rm miss}(B\pi)$ and $M(B)$ and to improve 
the resolution, we use 
$M_{\rm miss}^*=M_{\rm miss}(B\pi)+M(B)-m_B$ instead of 
$M_{\rm miss}(B\pi)$. The $M_{\rm miss}^*$ distribution for the RS 
combinations is shown in Fig.~\ref{fig:select}(b), where peaks 
corresponding to the $\bbstpi$ and $\bstbstpi$ signals are evident.
Combinations with $\pi^+$ --- the wrong sign (WS) combinations --- 
are used to evaluate the shape of the combinatorial background.
There is also a hint for a peaking structure in the WS $M_{\rm miss}^*$ 
distribution, shown as a hatched histogram in Fig.~\ref{fig:select}(b). 
Due to $B^0-\bar{B}^0$ oscillations, we expect a fraction of
the produced $B^0$ mesons to decay as $\bar{B}^0$ given by
$0.5x^2_d/(1+x^2_d)=0.1861\pm0.0024$, where $x_d$ is the $B^0$ 
mixing parameter~\cite{PDG}.


Note that the momentum spectrum of $B$ mesons 
produced in events with initial-state radiation (ISR), 
$\ee\to\gamma B\bar{B}$,
overlaps significantly with that for $B$ mesons from the three-body 
$\ee\to\bbpi$ processes. However, ISR events do not produce 
peaking structures in the $M_{\rm miss}^*$ distribution.

A binned maximum likelihood fit is performed to fit the 
$M_{\rm miss}^*$ distribution to the sum of three Gaussian functions to 
represent three possible signals and two threshold components 
$A_k(x_k-x)^{\alpha_k}\exp\{(x-x_k)/\delta_k\}$ ($k=1,2$) to parameterize
the $q\bar{q}$ and two-body $B^{(*)}\bar{B}^{(*)}$ backgrounds. 
The means and widths of the signal Gaussian functions are fixed from 
the signal MC simulation. The parameters $A_k$, $\alpha_k$, $\delta_k$ 
of the background functions are free parameters of the fit; 
the threshold parameters $x_k$ are fixed from the generic MC. 
ISR events produce an $M_{\rm miss}^*$ distribution similar to that 
for $q\bar{q}$ events; these two components are modeled by a single 
threshold function. 
The resolution of the signal peaks in Fig.~\ref{fig:select}(c) is 
dominated by the c.m.\ energy spread and is fixed at $6.5$~MeV/$c^2$ 
as determined from the signal MC. The fit to the RS spectrum 
yields $N_{BB\pi}=13\pm25$, $N_{\bbstpi}=357\pm30$ and 
$N_{\bstbstpi}=161\pm21$ signal events. The statistical significance of 
the observed  $\bbstpi$ and $\bstbstpi$ signal is $9.3\sigma$ and 
$8.1\sigma$, respectively. The statistical significance is 
calculated as $\sqrt{-2\ln({\cal L}_0/{\cal L}_{\rm sig})}$, where 
${\cal L}_{\rm sig}$ and ${\cal L}_0$ denote the likelihood values 
obtained with the nominal fit and with the signal yield fixed at zero,
respectively. 

\begin{figure}[!t]
\includegraphics[width=0.46\textwidth]{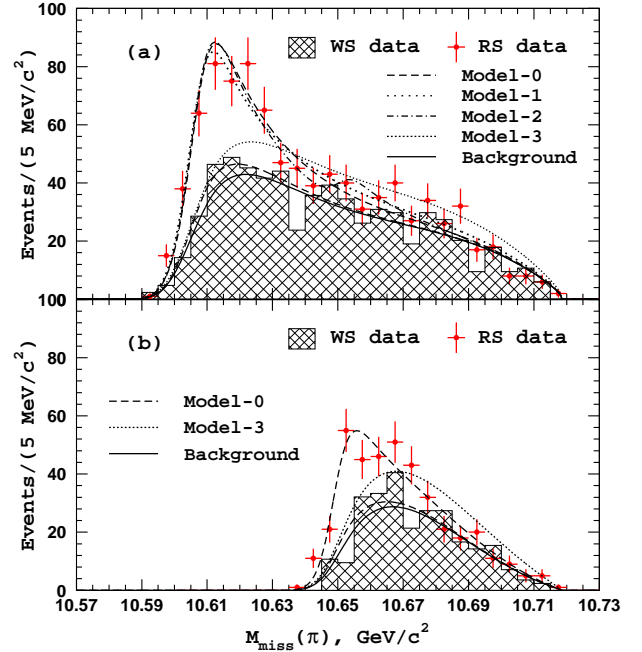} 
\hfill
  \caption{The $M_{\rm miss}(\pi)$ distribution for the 
           (a) $\bbstpi$ and (b) $\bstbstpi$ candidate events.}
  \label{fig:sig}
\end{figure}

For the subsequent analysis, we require 
$|M_{\rm miss}^*-m_{B^*}|<15$~MeV/$c^2$ to select $\bbstpi$ signal 
events and $|M_{\rm miss}^*-(m_{B^*}+\Delta m_B)|<12$~MeV/$c^2$, where 
$\Delta m_B=m_{B^*}-m_B$, to select $\bstbstpi$ events. For the 
selected $B^*B^{(*)}\pi$ candidates, we calculate 
$M_{\rm miss}(\pi) = \sqrt{(\sqrt{s}-E_\pi)^2/c^4-P^2_{\pi}/c^2}$, 
where $E_{\pi}$ and $P_{\pi}$ are the reconstructed energy and 
momentum, respectively, of the charged pion in the c.m.\ frame. 
The $M_{\rm miss}(\pi)$ distributions are shown in 
Fig.~\ref{fig:sig}~\cite{supl}. We perform a simultaneous binned maximum 
likelihood fit to the RS and WS samples, assuming the 
same number and distribution of background events in both samples and 
known fraction of signal events in the RS sample that leaks to the WS 
sample due to mixing. To fit the $M_{\rm miss}(\pi)$ spectrum, we use 
the function
\begin{eqnarray}
F(m) = [f_{\rm sig}S(m) + B(m)]\epsilon(m)F_{\rm PHSP}(m),
\label{eq:sigfit}
\end{eqnarray}
where $m \equiv M_{\rm miss}(\pi)$; $f_{\rm sig}=1.0$ 
($0.1105\pm0.0016$, \cite{fsig}) for the RS (WS) sample; $S(m)$ and 
$B(m)$ are the signal and background PDFs, respectively; and 
$F_{\rm PHSP}(m)$ is the phase space function. To account for the 
instrumental resolution, we smear the function $F(m)$ with a 
Gaussian function. The reconstruction efficiency is parametrized as
$\epsilon(m) \sim \exp((m-m_0)/\Delta)(1-m/m_0)^{3/4},$
where $m_0=10.718\pm0.001$~GeV/$c^2$ is an efficiency threshold and 
$\Delta=0.094\pm0.002$~GeV/$c^2$.


\begin{table*}[!t]
\centering
\caption{Summary of fit results to the $M_{\rm miss}(\pi)$ distributions 
         for the three-body $\bbstpi$ and $\bstbstpi$ final states.}
\medskip
\label{tab:results}
  \begin{tabular}{l|c|cccccc}  \hline \hline
 Mode ~~~& Parameter & 
~Model-0~ & \multicolumn{2}{c}{Model-1} & \multicolumn{2}{c}{Model-2} & ~Model-3~ \\
      & & & Solution 1  & Solution 2 & Solution 1  & Solution 2  \\
\hline 
$\bbstpi$
& $f_{\Zbl}$      &   $1.0$   &  $1.45\pm0.24$  & $0.64\pm0.15$ & $1.01\pm0.13$  & $1.18\pm0.15$ &  $-$ \\
& $f_{\Zbh}$      &   $-$     &       $-$       &      $-$      & $0.05\pm0.04$  & $0.24\pm0.11$ &  $-$\\
& $\phi_{\Zbh}$, rad.   &    $-$    &       $-$       &      $-$      & $-0.26\pm0.68$ & $-1.63\pm0.14$ & $-$ \\
& $f_{\rm nr}$    &    $-$    &  $0.48\pm0.23$  & $0.41\pm0.17$ &      $-$       &       $-$      & $1.0$\\
& $\phi_{\rm nr}$, rad. &   $-$     & $-1.21\pm0.19$  & $0.95\pm0.32$ &     $-$        &      $-$      &  $-$ \\
& $-2\log{\cal{L}}$& $-304.7$ &    $-300.6$    &    $-300.5$   &    $-301.4$    &     $-301.4$  & $-344.5$ \\
\hline
$\bstbstpi$
& $f_{\Zbh}$      &    $1.0$    & $1.04\pm0.15$ & $0.77\pm0.22$ & & &    $-$      \\
& $f_{\rm nr}$          &     $-$     & $0.02\pm0.04$ & $0.24\pm0.18$ & & &   $1.0$     \\
& $\phi_{\rm nr}$, rad. &     $-$     & $0.29\pm1.01$  & $1.10\pm0.44$  & & &   $-$      \\
& $-2\log{\cal{L}}$&  $-182.4$  &    $-182.4$    &    $-182.4$    & & &  $-209.7$    \\
\hline \hline
  \end{tabular}
\end{table*}

%
The distribution of background events is parameterized as 
$B_{B^{(*)}B^*\pi}(m) = b_0e^{-\beta\delta_m},$ where $b_0$ and $\beta$ 
are fit parameters and $\delta_m=m-(m_{B^{(*)}}+m_{B^*})$. A general 
form of the signal PDF is written as
\begin{eqnarray}
S(m)   = |{\cal{A}}_{Z_b(10610)} + {\cal{A}}_{Z_b(10650)} + {\cal{A}}_{\rm nr}|^2,
\label{eq:bstbstp-sign}
\end{eqnarray}
where ${\cal{A}}_{\rm nr}=a_{\rm nr}e^{i\phi_{\rm nr}}$ is the non-resonant 
amplitude parameterized as a complex constant and the two $Z_b$ 
amplitudes, ${\cal{A}}_{Z_b}$, are parameterized with Breit-Wigner 
functions ${\cal{A}}_{Z_b}=a_Ze^{i\phi_Z}/(m^2-m^2_Z-i\Gamma_Zm_Z)$. 
The masses and widths of 
the $Z_b$ states are fixed at the values obtained from the analyses 
of the $\ee\to\Upsilon(n{\rm S})\pp$ and $\ee\to h_b(m{\rm P})\pp$: 
$M_{\Zbl} = 10607.2\pm2.0$~MeV/$c^2$, $\Gamma_{\Zbl}=18.4\pm2.4$~MeV and
$M_{\Zbh} = 10652.2\pm1.5$~MeV/$c^2$, $\Gamma_{\Zbh}=11.5\pm2.2$~MeV
\cite{y5s2ypp}.

We first analyze of the $BB^*\pi$ [$B^*B^*\pi$] data with the simplest 
hypothesis, Model-0, that includes only the $\Zbl$ [$\Zbh$] amplitude. 
Results of the fit are shown in Fig.~\ref{fig:sig}; the numerical results 
are summarized in Table~\ref{tab:results}. The fraction $f_X$ of the 
total three-body signal attributed to a particular quasi-two-body 
intermediate state is calculated as
$ f_X = {\int |{\cal A}_X|^2\, dm}/{\int S(m)\, dm}$,
where ${\cal A}_X$ is the amplitude for a particular component $X$ of
the three-body amplitude. Next, we extend the hypothesis to include a 
possible non-resonant component, Model-1, and repeat the fit to the 
data. Then the $BB^*\pi$ data is fit to a combination of two $Z_b$ 
amplitudes, Model-2. In both cases, the addition of an extra component
to the amplitude does not give a statistically significant improvement 
in the data description: the likelihood value is only marginally improved 
(see Table~\ref{tab:results}). The addition of extra components to the 
amplitude also produces multiple maxima in the likelihood function. 
As a result, we use Model-0 as our nominal hypothesis.
Finally, we fit both samples to a pure non-resonant amplitude (Model-3). 
In this case, the fit is significantly worse.

If the parameters of the $Z_b$ resonances are allowed to float, the fit 
to the $\bbstpi$ data with Model-0 gives $10605\pm6$~MeV/$c^2$ and 
$25\pm7$~MeV for the $\Zbl$ mass and width, respectively, and the fit 
to the $B^*B^*\pi$ data gives $10648\pm13$~MeV/$c^2$ and $23\pm8$~MeV 
for the $\Zbh$ mass and width, respectively. The large errors here 
reflect the strong correlation between the resonance parameters.


The three-body visible~\cite{note1} cross sections are 
calculated as
\begin{eqnarray}
\sigma_{\rm vis}(\ee\to f) = 
\frac{N_{f}}
{L\cdot{\cal{B}}_{f}\cdot\alpha\cdot\eta},
\end{eqnarray}
where $N_{f}$ is the three-body signal yield and $L=121.4$~fb$^{-1}$ 
is the total integrated luminosity. 
The efficiency-weighted sum of $B$-meson branching fractions 
${\cal{B}}_{f}$ is determined using both signal MC and two-body 
$\ee\to B^{(*)}\bar{B}^{(*)}$ events in data. To avoid the large 
systematic uncertainties associated with the determination of 
reconstruction efficiencies for $B$ and $D$ decays to multibody 
final states, we select a subset of two-body modes: 
$B^+\to \bar{D}^0[K^+\pi^-]\pi^+$ and $B\to J/\psi[l^+l^-] K$, 
and calculate 
${\cal{B}}_{f}={\cal{B}}^{\rm\, sel}_{f}\times r\times
N^{\rm all}_{B^{(*)}\bar{B}^{(*)}}/
N^{\rm sel}_{B^{(*)}\bar{B}^{(*)}}$,
where the superscripts ``sel'' and ``all'' refer to quantities 
determined for the selected subset of $B$ decay modes and for the 
full set of modes, respectively. Two-body 
$\ee\to B^{(*)}\bar{B}^{(*)}$ events are selected with the requirement 
$0.90$~GeV/$c$ $<P(B)<1.35$~GeV/$c$; the $B$ yield is determined 
from the fit to the $M(B)$ distribution. We find 
$N^{\rm\, all}_{B^{(*)}\bar{B}^{(*)}}=10131\pm152$ and
$N^{\rm\, sel}_{B^{(*)}\bar{B}^{(*)}}=2406\pm62$.
To correct for a possible dependence of the $B$ meson reconstruction 
efficiency on its momentum, we calculate the double ratio $r$ 
using MC simulation:
%
$r=
   ({\cal{B}}^{\rm\, sel}_{B^{(*)}\bar{B}^{(*)}}/
    {\cal{B}}^{\rm\, all}_{B^{(*)}\bar{B}^{(*)}})/
   ({\cal{B}}^{\rm\, sel}_{B^{(*)}B^{(*)}\pi}/
    {\cal{B}}^{\rm\, all}_{B^{(*)}B^{(*)}\pi})=
1.0017\pm0.0096.$
To account for the non-uniform 
distribution of signal events over the phase space, we introduce an 
efficiency correction factor $\eta$ determined from the MC 
simulation with signal events generated according to the nominal 
model. Since we do not observe a signal in the $BB\pi$ final state, 
no correction is made for this channel. A factor $\alpha=0.897\pm0.007$
is introduced to correct for the effect of neutral $B$-meson 
oscillations that is determined using the known $B^0$ mixing parameter 
$x_d$ and the yield ratio in data of three-body events with a 
reconstructed neutral vs.\ charged $B$ meson.
The results are summarized in Table~\ref{tab:csec}.

\begin{table}[!t]
\centering
\caption{Summary of results on three-body cross sections.
         The first (or sole) uncertainty is 
         statistical; the second is systematic.}
\medskip
\label{tab:csec}
  \begin{tabular}{lccc}  \hline \hline
  ~Parameter~\hspace*{0mm} & $BB\pi$  & $BB^*\pi$ & $B^*B^*\pi$     \\
\hline 
Yield, Events  & $13\pm25$ & $357\pm30$ &  $161\pm21$  \\
${\cal{B}}_{f}$, $10^{-6}$ & $293\pm22$ & $276\pm21$ & $223\pm17$\\
$\eta$ & $1.0$ & $1.066$ & $1.182$\\
\hline
$\sigma_{\rm vis}$, pb   & $<2.1$ & $11.2\pm1.0\pm1.2$ & $5.61\pm0.73\pm0.66$ \\
\hline \hline
  \end{tabular}
\end{table}

The dominant sources of systematic uncertainties for the 
three-body production cross sections are the uncertainties 
in the signal yield extraction (6.9\% for $BB^*\pi$ 
and 8.7\% for $B^*B^*\pi$), 
in the reconstruction efficiency (7.6\%) (including secondary 
branching fractions~\cite{PDG}), 
in the correction factor $\alpha$ (1\%),
and the uncertainty in the integrated luminosity (1.4\%). 
The overall systematic uncertainties for the three-body cross sections
are estimated to be $7.9$\%, $10.4$\%, and $11.7$\% for the $BB\pi$, 
$\bbstpi$, and $\bstbstpi$ final states, respectively.

Using the results of the fit to the $M_{\rm miss}(\pi)$ spectra with 
the nominal model (Model-0 in Table~\ref{tab:results}) and the results 
of the analyses of $\ee\to\Upsilon(n{\rm S})\pp$~\cite{y5s2ypp} and 
$\ee\to h_b(m{\rm P})\pp$~\cite{y5s2hpp,note2}, we calculate the 
ratio of the branching fractions 
${\cal{B}}(\Zbl\to B\bar{B}^*+c.c.)/{\cal{B}}(\Zbl\to 
{\rm bottomonium}) = 4.76\pm0.64\pm0.75$ and 
${\cal{B}}(\Zbh\to B^*\bar{B}^*)/{\cal{B}}(\Zbh\to 
{\rm bottomonium}) = 2.40\pm0.44\pm0.50$.
%

We calculate the relative fractions for $Z_b$ decays, 
assuming that they are saturated by the already observed
$\Upsilon(n{\rm S})\pi$, $h_b(m{\rm P})\pi$, and $B^*B^{(*)}$ 
channels. The results are summarized in 
Table~\ref{tab:zfracs}.

In conclusion, we report the first observations of the three-body 
$\ee\to\bbstpi$ and $\ee\to\bstbstpi$ processes with a statistical
significance above $8\sigma$. Measured visible cross sections are
$\sigma_{\rm vis}(\ee\to\bbstpi)=(11.2\pm1.0\pm1.2)$~pb and 
$\sigma_{\rm vis}(\ee\to\bstbstpi=(5.61\pm0.73\pm0.66)$~pb. For the 
$\ee\to BB\pi$ process, we set a 90\% confidence level upper limit 
of $\sigma_{\rm vis}(\ee\to BB\pi)<2.1$~pb. The analysis of the 
$B^{(*)}B^*$ mass spectra indicates that the total three-body rates 
are dominated by the intermediate $\ee\to\Zbl^\mp\pi^\pm$ and 
$\ee\to\Zbh^\mp\pi^\pm$ transitions for the $\bbstpi$ and $\bstbstpi$ 
final states, respectively.


We thank the KEKB group for excellent operation of the accelerator; 
the KEK cryogenics group for efficient solenoid operations; and the KEK
computer group, the NII, and PNNL/EMSL for valuable computing and SINET4 
network support. We acknowledge support from MEXT, JSPS and Nagoya's 
TLPRC (Japan); ARC and DIISR (Australia); FWF (Austria); NSFC (China); 
MSMT (Czechia); CZF, DFG, and VS (Germany); DST (India); INFN (Italy); 
MEST, NRF, GSDC of KISTI, and WCU (Korea); MNiSW and NCN (Poland); 
MES and RFAAE (Russia); ARRS (Slovenia); IKERBASQUE and UPV/EHU (Spain); 
SNSF (Switzerland); NSC and MOE (Taiwan); and DOE and NSF (USA).

\begin{table}[!t]
\centering
\caption{$B$ branching fractions for the $Z^+_b(10610)$ and 
         $Z^+_b(10650)$ decays. The first quoted uncertainty is 
         statistical; the second is systematic.}
\medskip
\label{tab:zfracs}
  \begin{tabular}{lcc}  \hline \hline
  ~Channel~\hspace*{5mm}  & \multicolumn{2}{c}{Fraction, \%}   \\
              & ~~~~~~~~$\Zbl$~~~~~~~~  & ~~$\Zbh$~~      \\
\hline 
 $\Upsilon(1{\rm S})\pi^+$ & $0.60\pm0.17\pm0.07$ & $0.17\pm0.06\pm0.02$ \\
 $\Upsilon(2{\rm S})\pi^+$ & $4.05\pm0.81\pm0.58$ & $1.38\pm0.45\pm0.21$ \\
 $\Upsilon(3{\rm S})\pi^+$ & $2.40\pm0.58\pm0.36$ & $1.62\pm0.50\pm0.24$ \\
 $h_b(1{\rm P})\pi^+$     & $4.26\pm1.28\pm1.10$ & $9.23\pm2.88\pm2.28$ \\
 $h_b(2{\rm P})\pi^+$     & $6.08\pm2.15\pm1.63$ & $17.0\pm3.74\pm4.1$  \\
 $B^+\bar{B}^{*0}+\bar{B}^0B^{*+}$
                          &  $82.6\pm2.9\pm2.3$  &         $-$         \\
 $B^{*+}\bar{B}^{*0}$      &         $-$          & $70.6\pm4.9\pm4.4$  \\
\hline \hline
  \end{tabular}
\end{table}



\begin{thebibliography}{99}
\bibitem{note1} We do not apply an ISR correction 
here since the energy dependence of the cross section is not known.
\end{thebibliography}
\end{document}


\title{Supplementary Material}
\maketitle

\section{$M_{\rm miss}(\pi)$ Spectra}

In this supplementary material, we present the bin-by-bin
information for the $M_{\rm miss}(\pi)$ distribution shown 
in Fig.~2 of the Letter.
Results are summarized in Table~\ref{tab:corr}, where the 
quoted uncertainty is statistical only. In addition to the 
right and wrong-sign yields, the detector relative efficiency 
is presented.

\begin{table*}[!hb]
\centering
\caption{The three-body $B^{(*)}B^*\pi$ signal yields in $M_{\rm miss}(\pi)$ bins.}
\medskip
\label{tab:corr}
  \begin{tabular}{l|ccc|ccc}  \hline \hline
  $M_{\rm miss}(\pi)$ bin, MeV/$c^2$  
   &  \multicolumn{3}{c|}{$BB^*\pi$} & \multicolumn{3}{c}{$B^*B^*\pi$}   \\ 
   & RS data, & WS data, & Relative efficiency, & RS data, & WS data, & Relative efficiency, \\
   & Events   & Events   & Arbitrary units      & Events   & Events   & Arbitrary units \\
\hline 
 $10590-10595$ & $ 1\pm1.0$ & $ 2.3\pm1.7$ & $0.937$ & $-$ & $-$ & $-$ \\                
 $10595-10600$ & $15\pm3.9$ & $ 4.7\pm2.4$ & $0.958$ & $-$ & $-$ & $-$ \\                
 $10600-10605$ & $38\pm6.2$ & $14.3\pm4.1$ & $0.979$ & $-$ & $-$ & $-$ \\                
 $10605-10610$ & $64\pm8.0$ & $28.6\pm5.8$ & $0.999$ & $-$ & $-$ & $-$ \\                
 $10610-10615$ & $81\pm9.0$ & $46.4\pm7.4$ & $1.017$ & $-$ & $-$ & $-$ \\                
 $10615-10620$ & $75\pm8.7$ & $48.8\pm7.6$ & $1.034$ & $-$ & $-$ & $-$ \\                
 $10620-10625$ & $81\pm9.0$ & $45.2\pm7.3$ & $1.050$ & $-$ & $-$ & $-$ \\                
 $10625-10630$ & $65\pm8.1$ & $41.7\pm7.0$ & $1.064$ & $-$ & $-$ & $-$ \\                
 $10630-10635$ & $47\pm6.9$ & $44.1\pm7.2$ & $1.075$ & $-$ & $-$ & $-$ \\
 $10635-10640$ & $45\pm6.7$ & $23.8\pm5.3$ & $1.084$ & $ 1\pm1.0$ & $ 0.0\pm1.1$ & $1.084$ \\
 $10640-10645$ & $39\pm6.2$ & $35.7\pm6.5$ & $1.089$ & $11\pm3.3$ & $ 0.0\pm1.1$ & $1.089$ \\
 $10645-10650$ & $43\pm6.6$ & $39.3\pm6.8$ & $1.091$ & $21\pm4.6$ & $10.8\pm3.6$ & $1.091$ \\
 $10650-10655$ & $40\pm6.3$ & $34.5\pm6.4$ & $1.089$ & $55\pm7.4$ & $ 9.6\pm3.4$ & $1.089$ \\
 $10655-10660$ & $31\pm5.6$ & $26.2\pm5.6$ & $1.082$ & $45\pm6.7$ & $32.2\pm6.2$ & $1.082$ \\
 $10660-10665$ & $35\pm5.9$ & $30.9\pm6.1$ & $1.070$ & $46\pm6.8$ & $33.4\pm6.3$ & $1.070$ \\
 $10665-10670$ & $40\pm6.3$ & $29.8\pm5.9$ & $1.051$ & $51\pm7.1$ & $40.5\pm6.9$ & $1.051$ \\
 $10670-10675$ & $27\pm5.2$ & $19.0\pm4.7$ & $1.025$ & $43\pm6.6$ & $21.4\pm5.0$ & $1.025$ \\
 $10675-10680$ & $34\pm5.8$ & $29.8\pm5.9$ & $0.991$ & $32\pm5.7$ & $27.5\pm5.7$ & $0.991$ \\
 $10680-10685$ & $26\pm5.1$ & $27.4\pm5.7$ & $0.946$ & $21\pm4.6$ & $16.7\pm4.4$ & $0.946$ \\
 $10685-10690$ & $32\pm5.7$ & $20.2\pm4.9$ & $0.891$ & $18\pm4.2$ & $16.7\pm4.4$ & $0.891$ \\
 $10690-10695$ & $17\pm4.1$ & $ 9.5\pm3.4$ & $0.821$ & $20\pm4.5$ & $14.3\pm4.1$ & $0.821$ \\
 $10695-10700$ & $18\pm4.2$ & $17.8\pm4.6$ & $0.735$ & $11\pm3.3$ & $15.5\pm4.3$ & $0.735$ \\
 $10700-10705$ & $ 8\pm2.8$ & $ 9.5\pm3.4$ & $0.629$ & $ 9\pm3.0$ & $ 7.1\pm2.9$ & $0.629$ \\
 $10705-10710$ & $ 8\pm2.8$ & $10.7\pm3.6$ & $0.495$ & $ 5\pm2.2$ & $ 3.6\pm2.1$ & $0.495$ \\
 $10710-10715$ & $ 6\pm2.4$ & $ 5.9\pm2.7$ & $0.322$ & $ 5\pm2.2$ & $ 2.4\pm1.7$ & $0.322$ \\
 $10715-10720$ & $ 2\pm1.4$ & $ 0.0\pm1.1$ & $0.056$ & $ 1\pm1.0$ & $ 0.0\pm1.1$ & $0.056$ \\
\hline \hline
  \end{tabular}
\end{table*}